\documentclass{emulateapj}

\submitted{To Appear in the Astrophysical Journal Letters}
\shortauthors{ODA ET AL.}
\shorttitle{Gamma-rays from First Cosmological Objects}

\begin{document}

\title{Gamma-Ray Background from Neutralino Annihilation in the First
Cosmological Objects}

\author{TAKESHI ODA \altaffilmark{1}, 
TOMONORI TOTANI \altaffilmark{1} AND MASAHIRO
NAGASHIMA \altaffilmark{2}}

\altaffiltext{1}{Department of Astronomy, School of Science, Kyoto
University, Sakyo-ku, Kyoto 606-8502, Japan}
\altaffiltext{2}{Department of Physics, School of Science, Kyoto
University, Sakyo-ku, Kyoto 606-8502, Japan}

\email{takeshi@kusastro.kyoto-u.ac.jp}

\begin{abstract}
The paradigm of the neutralino dark matter predicts that the first
gravitationally bound objects are earth-mass sized microhaloes, which
would emit annihilation gamma-rays. Here we show that, though the flux
from individual nearest microhaloes is extremely difficult to detect,
meaningful constraints on their survival probability and internal
density profile can be set by requiring that the galactic and
extragalactic gamma-ray background flux from the microhaloes does not
exceed the existing EGRET background data.  Possible disruption of
microhaloes by stellar encounters does not significantly reduce the
background flux.  If the probability for microhaloes to survive the
hierarchical clustering process of dark matter is as large as indicated
by a recent simulation, they could be a significant component of the
observed background flux in some photon energy range, even with the
standard annihilation cross section and conservative internal density
profile of microhaloes.  The integrated gamma-ray flux from microhaloes
in the halo of the Andromeda galaxy may also be detectable by 
observations in the near future.
\end{abstract}

\keywords
{dark matter --- gamma rays: theory --- Galaxy: halo --- galaxies: haloes}

\section{INTRODUCTION}
The neutralino predicted by the supersymmetry theory of particle physics
is the most promising candidate of the cold dark matter (CDM).
The theory predicts that neutralinos
should annihilate and produce high energy particles such as gamma-rays,
and the detectability of these particles from dense regions like the Galactic
center (GC) has been discussed intensively in the literature. [See
\citet{bertone_etal04} for a review and references therein for the
earlier papers.] The structure formation theory predicts hierarchical
substructure in a dark matter halo, and this clumpiness is expected to
enhance the annihilation signal. Previous investigations, however,
considered substructures only to the first order (i.e., isolated
subhaloes in a halo) and masses larger than $\sim 10^6 M_\odot$,
mainly because of the limitation of cosmological N-body simulations
\citep{ullio_etal02,taylor_silk03,elsaesser_mannheim05}.

Recently attention has been turned to substructures on much smaller
scale, especially the first gravitationally bound objects in the
cosmological evolution. If the dark matter is the neutralino, any
density fluctuations of mass scales smaller than $M_{\rm mh} \sim
10^{-6} M_\odot$ are washed out by collisional damping and subsequent
free streaming in the early universe. Then the first objects form as
``microhaloes'' with a mass $\sim M_{\rm mh}$ at $z \sim z_{\rm nl}$,
where $z_{\rm nl} = 60 \pm 20$ is the epoch when the rms linear density
fluctuation at this mass scale, $\sigma(M_{\rm mh})$, becomes unity
\citep{hofmann_etal01,green_etal04,green_etal05,berezinsky_etal03,
loeb_zaldarriaga05}.  Therefore a considerable part of the neutralino
dark matter should have collapsed into these earth-mass objects, at
least once in the cosmic history.

However, it is highly uncertain how much fraction of these microhaloes
can survive the subsequent hierarchical structure formation until present. 
\citet{berezinsky_etal03} estimated that only
0.1--0.5 \% of microhaloes survive, because of tidal disruption when
they are taken into larger haloes. This is, however, a completely
analytic estimate and uncertainty must be large. On the other hand,
\citet{diemand_etal05}, based on a N-body simulation, argued that about
50\% of the total halo mass $M_{\rm tot}$ is in the form of substructure
with a subhalo mass function $dN/dM \propto M^{-\mu}$ in the mass range
$10^{-6} < M < 10^{10} M_\odot$, where $\mu \sim 2$. This indicates that
the mass fraction of microhaloes is at least $M_{\rm mh}^2 [dN(M_{\rm
mh})/dM] / M_{\rm tot} \sim 1.3\%$. Furthermore, the nested nature of
hierarchical structure formation predicts that microhaloes may be
embedded in larger subhaloes, which may again be embedded in even larger
ones. Counting microhaloes in larger mass subhaloes up to $M \sim
10^{10} M_\odot$ will further increase the true number of earth-mass
microhaloes. \citet{diemand_etal05} used $500 \ \rm pc^{-3}$ as the
number density of such microhaloes in the solar neighborhood, which is
about 7\% of the standard dark matter density at the Sun's location,
$\rho(R_\odot) = 0.3 \ \rm GeV \ cm^{-3}$ \citep{bertone_etal04}. 
In addition to the tidal disruption by hierarchical clustering, 
microhaloes may also be destroyed by tidal
interaction with stars in the Galactic disk
\citep{zhao_etal05,moore_etal05}, but again estimates are controversial.

It is obvious that such microhaloes could have significant impact on the
detectability of the annihilation signal. Here we show that both the
galactic and extragalactic gamma-ray background radiation (hereafter
GGRB and EGRB, respectively) give a strong constraint on the existence
of such microhaloes.  Since the microhalo survival probability is
highly uncertain, we simply parametrize this quantity as $f_{\rm surv}$,
and try to estimate how much constraints can be set from existing and
future observations.  It has been argued that neutralino annihilation
cannot be a significant component of the observed EGRB data  
since it would overpredict the
gamma-ray flux from the GC beyond the observational upper bound
\citep{ando05}. This argument does not apply here, because the
microhaloes are likely disrupted by strong tidal forces of the GC
gravity field and/or interaction with stars within $\sim$ kpc of the GC
\citep{diemand_etal05} and hence the visibility of the GC is not
enhanced by the microhaloes. However, the mass included within 1 (10)
kpc is only 0.3 (6) \% of the total mass of the Galactic halo
\citep{klypin_etal02}. Therefore the total gamma-ray flux from
microhaloes in a galactic halo, to which the EGRB is related, is hardly
affected by tidal disruption in its central region, if microhaloes trace
the mass distribution.

\section{Gamma-ray background from microhaloes}

\subsection{Estimating GGRB and EGRB Flux}

In this work we assume that microhaloes are formed at redshift $\sim z_{\rm
nl}$, and a fraction ($1- f_{\rm surv}$) of them are immediately
destroyed by tidal forces in subsequent hierarchical structure
formation. After that, we assume that the number of microhaloes and
their density profile are kept constant until present, except in regions
very close to galactic halo centers. These are not unreasonable, since the
tidal force within an object is proportional to its internal density, and the
mean density within virialized objects decreases with the cosmic expansion
in proportion to the mean background density. Therefore we expect that
the tidal disruption at a typical location within a dark halo should occur
most efficiently at redshift not very different from $z_{\rm nl}$, when
isolated microhaloes are first taken into larger objects.

Consider a region in the universe with the mass scale $M_{\rm mh}$ whose
linear fractional overdensity is $\delta = \delta \rho / \rho \propto
(1+z)^{-1}$. According to the standard structure formation theory, the
height over the rms, $\nu \equiv \delta / \sigma(M_{\rm mh})$, obeys to
the Gaussian, i.e., the comoving number density of microhaloes given by
$dn_{\rm mh}/d\nu = f_{\rm surv} \Omega_\chi \rho_{\rm crit, 0}
\exp(-\nu^2/2)/(\sqrt{2\pi} M_{\rm mh})$, where we use the WMAP values
\citep{spergel_etal03} for the present-day critical density $\rho_{\rm
crit, 0}$ and the neutralino density parameter $\Omega_\chi = 0.22$.  At
$z \sim z_{\rm nl}$ the universe is flat and matter-dominated, and the
region will collapse and virialize when $\delta$ grows to $\delta_c =
1.686$ at $z = z_{\rm vir}$, where $(1 + z_{\rm vir}) = (1+z_{\rm nl})
\nu / \delta_c$. The internal density of microhaloes is given as
$\rho_{\rm eff}(\nu) = 18 \pi^2 f_c (1+z_{\rm vir})^3 \Omega_\chi
\rho_{\rm crit, 0}$. Here, $f_c$ is the enhancement factor from the
virial density, to take into account the density profile of each
microhalo; we found that the mass-weighted mean density, which is
proportional to the annihilation rate, is increased by $f_c = 6.2$ for
the microhalo profile\footnote{
The $\alpha \beta \gamma$-profile with ($\alpha$, $\beta$, $\gamma$) =
(1, 3, 1.2) and the concentration parameter $c = 1.6$.} 
found in the simulation \citep{diemand_etal05}.

Then, integrating over $\nu$, the
comoving annihilation rate density is given by:
\begin{eqnarray}
\dot{{\cal N}}_{\chi \chi} &=&
\int_{\nu_l}^\infty M_{\rm mh}
\ \rho_{\rm eff}(\nu) \
\frac{
\langle \sigma_{\chi \chi} \upsilon \rangle }{2 m_\chi^2} \
\frac{dn_{\rm mh}}{d\nu} \ d\nu \ ,
\end{eqnarray}
where $m_\chi$ is the neutralino mass and $\langle
\sigma_{\chi \chi} \upsilon \rangle$ being the mean velocity-multiplied
cross
section of neutralino annihilation. Throughout this letter we use the
standard value of $\langle \sigma_{\chi\chi} \upsilon \rangle = 3 \times
10^{-26} \ \rm cm^3 s^{-1}$ \citep{bertone_etal04}, and scaling for
different values is obvious. The possible range of $m_\chi$
is $\sim 30$ GeV -- 10 TeV \citep{bertone_etal04}, and we use $m_\chi =
100$ GeV for calculations below, unless otherwise stated.
Since $\rho_{\rm eff} \propto \nu^3$,
the annihilation signal comes mainly from microhaloes of
$\sim \nu_p$ sigma fluctuation, where
$\nu_p \equiv \sqrt{3}$. Microhaloes with
$\nu \ll 1$ may not gravitationally collapse or
would be disrupted by subsequent structure formation, but
the integration is not very sensitive to the lower bound, and hence
we take $\nu_l = 0$. Note that, by this formulation, $f_{\rm surv}$
is effectively the survival probability for microhaloes having
relatively high density fluctuation of $\nu \sim \nu_p$, and their
mass fraction in the total dark matter, $f_m$, is related as:
$f_m \sim (1/\sqrt{2\pi}) \exp(-\nu_p^2/2) f_{\rm surv}
\sim 0.09 f_{\rm surv}$.
Assuming that the spatial distribution of microhaloes traces the
smoothed mass over larger scales, the number density of microhaloes with
$\nu \sim \nu_p$ around the solar system is $n(R_\odot) \sim
f_m \rho(R_\odot) / M_{\rm mh} \sim 680 f_{\rm surv} \ \rm pc^{-3}$.
This number is close to the estimate based on the simulation, $\sim
500 \ \rm pc^{-3}$ \citep{diemand_etal05}, indicating that $f_{\rm surv}$
could be of order unity.

Now we can calculate the EGRB photon flux from microhaloes per
steradian, as:
\begin{eqnarray}
\frac{dF_\gamma}{dE_\gamma} &=& \frac{c \ \dot{{\cal N}}_{\chi \chi}}{4\pi}
\int_0^{z_{\rm nl}} dz \frac{dt}{dz}
\frac{dn_\gamma[(1+z)E_\gamma]}{dE_\gamma} (1+z) \ ,
\end{eqnarray}
where $E_\gamma$ is the gamma-ray energy and its spectrum produced by an
annihilation, $dn_\gamma/dE_\gamma$, is calculated using the analytical
fitting formula given in \citet{bergstrom_etal01}. (We consider only the
continuum gamma-rays.) The integration up to the redshift $z_{\rm nl}$
is an approximation, but it is almost insensitive to this upper bound
since annihilation at $z \lesssim 1$ is dominant to the EGRB.  High
energy gamma-rays may be absorbed by interaction with the cosmic
infrared background. However, in this letter we consider only $E_\gamma
\leq 100$ GeV, and at this photon energy the optical depth becomes unity
only beyond $z \sim 2$ \citep{totani_takeuchi02}. Therefore our result
is hardly affected by the absorption.

Next we calculate the GGRB flux. 
The annihilation rate per unit dark matter mass in
a region smoothed over larger scales than microhaloes is given by
$\dot{N}_{\chi \chi} = \dot{\cal N}_{\chi \chi} / (\Omega_\chi \rho_{\rm
crit, 0})$. Then we obtain the GGRB flux per steradian as:
\begin{eqnarray}
\frac{dF_\gamma}{dE_\gamma} = \frac{\dot{N}_{\chi \chi}}{4\pi}
\frac{dn_\gamma(E_\gamma)}{dE_\gamma}
\int_{\rm l.o.s.} \rho_{\rm sm} \ dl \ ,
\end{eqnarray}
where the integration is over the line of sight, for the smoothed dark
matter density in the Galactic halo, $\rho_{\rm sm}$. It should be
noted that the flux is proportional to the line-of-sight integration of
$\rho_{\rm sm}^1$, {\it not} $\rho_{\rm sm}^2$ as in the case of diffuse
matter distribution. We use two spherically symmetric models of
$\rho_{\rm sm}$ \citep{klypin_etal02}; one has the Navarro-Frenk-White
\citep[NFW,][]{NFW} profile but the other is modified from the NFW profile
by adiabatic compression of dark matter responding to the baryon infall.
As discussed above, microhaloes are expected to be destroyed by tidal
forces in the inner region around the GC. Therefore, as a simple model,
we introduce the disruption radius $R_{d}$ within which microhaloes are
completely disrupted, while they are all preserved outside $R_d$ with an
abundance proportional to $f_{\rm surv}$.

\subsection{Comparison to Observations}

\begin{figure*}
\epsscale{0.6} \plotone{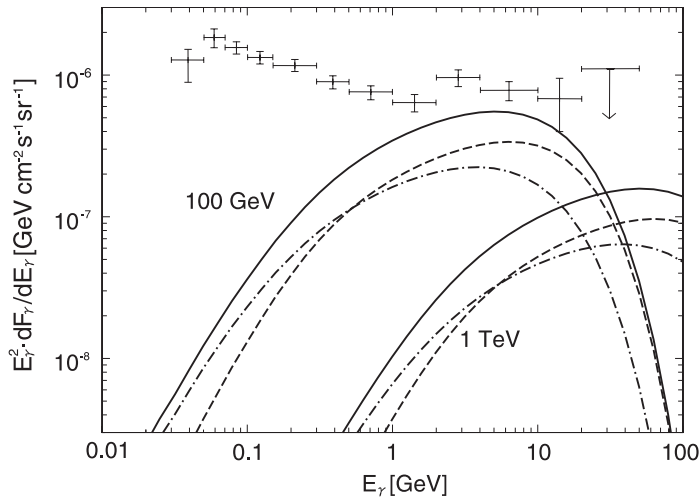} \caption{The background gamma-ray flux
from neutralino annihilation in the microhaloes. The Galactic (dashed),
extragalactic (dot-dashed), and the total (solid) components are
shown. The two cases of $m_\chi = $ 100 GeV and 1 TeV are presented,
with $f_{\rm surv} =$ 0.35 and 1, respectively.  The internal density
profile parameter $f_c = 6.2$ is conservatively assumed. The baryon
compressed NFW profile for the Galactic halo and $R_d = 5$ kpc are used
for the Galactic component.  
\label{fig:spec}}
\end{figure*}

The calculated EGRB flux is shown in Fig. \ref{fig:spec} for $m_\chi = $
100 GeV and 1 TeV. If $f_{\rm surv} \gtrsim 0.1$, the microhaloes make a
significant contribution to the observed EGRB flux
\citep{strong_etal04_egrb} in some photon energy range. The predicted
flux is much higher than those in earlier studies
\citep{ullio_etal02,taylor_silk03,ando05}; this is mainly because the
microhaloes formed much earlier than galactic haloes considered in these
studies ($M \gtrsim 10^{5-6} M_\odot$) and hence have much higher
internal density. The GGRB flux from the microhaloes, which is the mean
in the all sky except for the disk region \footnote{The removed disk
region is the same as defined by \cite{strong_etal04_egrb}, for a
consistent comparison between the prediction and the data.}, is also
shown in this figure.  It is found to be comparable to the EGRB,
indicating that it could also be a significant component of the observed
flux.  It should be noted that, since the ``observed'' EGRB data was
estimated as the residual after the subtraction of the cosmic-ray
interaction model in our Galaxy, not only the EGRB but also the GGRB
expected from microhaloes should be compared to the observed EGRB data.

\begin{figure*}
\epsscale{0.6}
\plotone{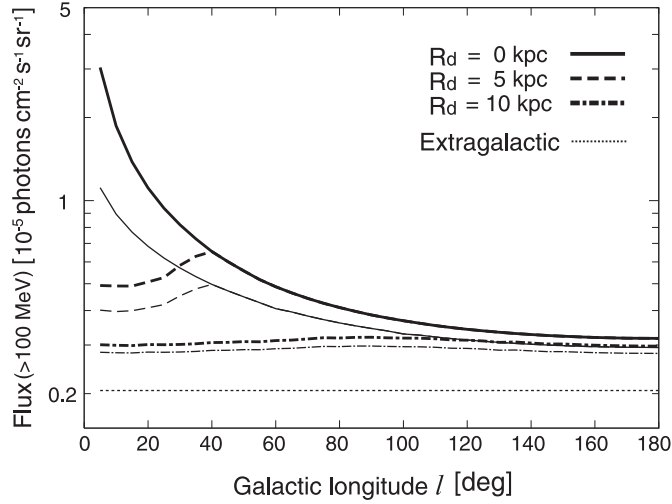}
\caption{The longitudinal distribution of the background flux
(GGRB+EGRB), assuming $m_\chi = 100$ GeV,
$f_{\rm surv} = 1$, and $f_c = 6.2$.
Three different values of disruption radius $R_d$ are
used as indicated. Two different density profiles of the Galactic
halo are used: the baryon-compressed NFW (upper thick curves)
and the original NFW (lower thin curves). The level of the 
predicted isotropic EGRB from microhaloes is
also indicated.
\label{fig:flux_l}}
\end{figure*}

Figure \ref{fig:flux_l} shows the predicted
total flux (EGRB $+$ GGRB) as a function of the Galactic longitude. It
can be seen that, if $R_d \gtrsim 5$ kpc, the anisotropy of the
background flux is at most a factor of 2. This is acceptable,
considering the precision of the EGRB flux measurements and possible
systematic uncertainties in the foreground subtraction
\citep{sreekumar_etal98,strong_etal04_egrb,keshet_etal04}. Even smaller
$R_d$ may also be allowed, since the anisotropy close to the GC would be
hidden by strong background flux from cosmic-ray interactions in the
Galactic disk. 

In fact, evidence for a diffuse gamma-ray halo towards
the GC that cannot be explained by the standard cosmic-ray interaction
model has been reported \citep{dixon_etal98},
at a flux level similar to the EGRB. This
gamma-ray halo might be explained by the microhaloes with an appropriate
choice of $R_d$. The GGRB sky distribution is expected to show strong
small-scale anisotropy by the complicated substructures in the Galactic
halo. Since annihilation signal peaks in rather narrow photon energy
range, the gamma-ray energy dependence of the GGRB anisotropy may be
used to examine the contribution from microhaloes.

The EGRET background data around the disk region shows so-called GeV
excess over the standard prediction from cosmic-ray interaction, which
is about 1--2 orders of magnitude higher than the EGRB along the
Galactic disk \citep{hunter_etal97}. The flux level of the excess might
be achieved by microhaloes with a large boost factor from our prediction
above, which is, in fact, not very unlikely (see below). However, still
the GeV excess seems difficult to explain by the microhaloes, because
the spatial distribution of the GeV excess is clearly associated with
the Galactic disk while distribution of microhaloes is expected to be
more spherical.  Note that the GeV excess can also be explained by
modification of the cosmic-ray interaction models
\citep{strong_etal04,kamae_etal05}.

It should be noted that the internal density profile of the microhaloes used
above is conservative in a sense that it predicts relatively low
annihilation luminosity. Though we have used the $\alpha \beta \gamma$
profile with $\gamma=1.2$ following the fitting by
\citet{diemand_etal05} ($\rho \propto r^{-\gamma}$ with $r
\rightarrow 0$), their simulated microhaloes have $\gamma \sim 1.7$ to
the resolution limit of the simulation. They also noticed the
similarity between their simulated microhaloes and galactic haloes
shortly after the formation or major mergers,
showing a single power law profile with
slopes of $\gamma \sim 1.5$--2. If $\gamma > 1.5$, annihilation
luminosity diverges in the center, and assuming that the maximum density
is limited by annihilation time scale, $\rho_{\max} \langle \sigma_{\chi
\chi} \upsilon \rangle / (2 m_\chi) < (10^{10}$ yr)$^{-1}$, we find that
the enhancement factor $f_c = 6.2$ used above (for $\gamma = 1.2$) will
be boosted up to $f_c = $ 31, 190, and $1.4 \times 10^4$ for $\gamma =
1.5, 1.7$, and 2.0, respectively. Here we used consistently 
(and conservatively) a low value
for the concentration parameter, $c= 1.6$, as found in the simulation
\citep{diemand_etal05}.

\begin{figure*}
\epsscale{0.55} \plotone{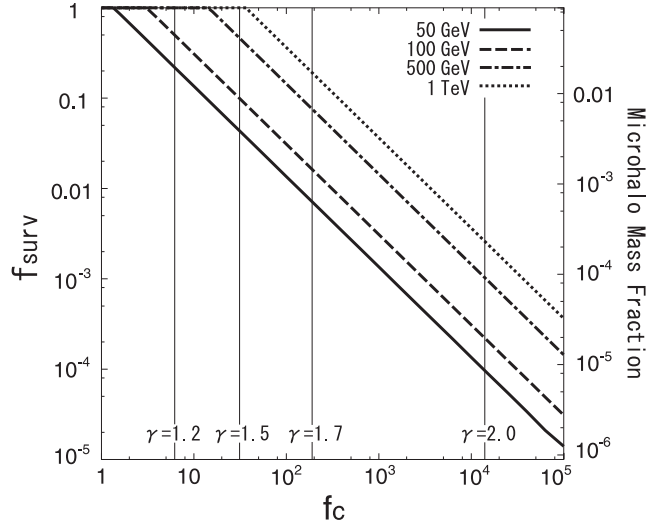} \caption{The excluded region in the
space of the two parameters, the survival probability ($f_{\rm surv}$)
and the enhancement factor by the internal density profile of
microhaloes ($f_c$). The microhalo mass fraction $f_m$ in the total dark
matter mass is related to $f_{\rm surv}$ as $f_m \sim 0.09 f_{\rm
surv}$. Several curves are depicted for different neutralino masses
$(m_\chi)$ as indicated, and the upper-right regions are excluded
because the predicted background flux will exceed the observed data. The
compressed NFW profile for the Milky Way halo and $R_d = 5$ kpc are
assumed for the GGRB component. The values of $f_c$ corresponding to
several values of the inner slope index $(\gamma)$ of
internal density profile of microhaloes are marked by vertical thin solid
lines. \label{fig:f_surv}}
\end{figure*}

Provided that annihilation cross section is close to the standard value,
the two major astrophysically uncertain parameters are $f_{\rm surv}$
and $f_c$. We note that the background flux is only weakly dependent on
$M_{\rm mh}$. The formation redshift $z_{\rm nl}$ depends 
on $M_{\rm mh}$, but only very weakly, 
because $\sigma(M)$ is only weakly dependent on
$M$ in small scales under the standard CDM density fluctuation spectrum
and hence fluctuations over a wide range of mass scales will become
non-linear at similar redshifts.  Hence we calculate the excluded region
in the $f_{\rm surv}$-$f_c$ plane in Fig. \ref{fig:f_surv}, by requiring
that the predicted flux does not exceed any of the observed EGRB data.

\section{Discussion and Conclusions}
The observability of the nearest microhalo from the Earth is of
great interest. The expected photon flux is given by:
\begin{eqnarray}
F &\sim& \frac{ Y_\gamma M_{\rm mh} \rho_{\rm eff}
}{4 \pi \ [n(R_\odot)]^{-2/3}}
\frac{\langle \sigma_{\chi \chi} \upsilon \rangle}
{2 m_\chi^2} \\
&=& 1.5 \times 10^{-11} f_{\rm surv}^{2/3}
(f_c/6.2) Y_{40} m_{2}^{-2}
\ \rm cm^{-2} s^{-1} \ ,
\end{eqnarray}
where $m_{2} = m_\chi$ / (100 GeV) and $Y_\gamma \equiv 40 Y_{40}$
is the photon number yield from one annihilation; typically one
annihilation produces 30--50 continuum gamma-rays and more than 80\% of
them are above 100 MeV \citep{darksusy}. Unfortunately this flux is much
smaller than the point source sensitivity of the EGRET, and even of the
future GLAST mission, and hence detection of individual microhaloes is
unlikely. It is impossible to greatly enhance the detectability by
boosting up the internal density factor $f_c$, because it would
seriously overpredict the background flux far beyond the observed level.

The detectability of gamma-rays from microhaloes in the nearby
extragalactic objects is also intriguing. Here we estimate the flux
expected from M31, $d = $ 770 kpc from the Earth. The dark matter mass
enclosed within 13 kpc, corresponding to the position accuracy of
EGRET $(\sim 1^\circ)$, is $M_{DM} \sim 1.5 \times 10^{11} M_\odot$
\citep{klypin_etal02}. Then we expect photon flux of $Y_\gamma
\dot{N}_{\chi \chi} M_{DM} / (4 \pi d^2) \sim 7.7 \times 10^{-9} f_{\rm
surv} (f_c/6.2)
Y_{40} m_{2}^{-2} \ \rm cm^{-2} s^{-1}$, which is interestingly
very close to the EGRET upper bound \citep[$1.6 \times 10^{-8}$,][]
{blom_etal99}, indicating that there is a good chance of detecting
annihilation flux from microhaloes in M31 by the GLAST mission or next
generation air Cerenkov telescopes (ACTs), with a larger $m_\chi$
favoring the latter. In contrast to the flux from the center of cuspy
density profile, we expect rather diffuse flux distribution around the
center of M31, because microhaloes that are very close to the M31 center
are likely disrupted. The improved angular resolution of the GLAST or
ACTs will be able to resolve it.

In conclusion, annihilation gamma-rays from the microhaloes could be a
significant component of the observed EGRB flux. The two major uncertain
parameters are $f_{\rm surv}$ and $f_c$, and a considerable part of the
parameter space has already been excluded by the observed EGRB flux
level. 

This work was supported by the Grant-in-Aid for the 21st Century COE
"Center for Diversity and Universality in Physics" from the Ministry of
Education, Culture, Sports, Science and Technology (MEXT) of Japan.



\end{document}